%% file: pr.tex
\documentstyle[sprocl,epic,eepic,epsfig]{article}

\def\be{\begin{equation}}
\def\ee{\end{equation}}
\def\bea{\begin{eqnarray}}
\def\eea{\end{eqnarray}}
\def\Sp{\mbox{Sp}}

\def\Re{\mbox{Re}}

\begin{document}

\title{DYNAMICAL SYMMETRY BREAKING IN FOUR-FERMIONIC MODELS UNDER THE
INFLUENCE OF EXTERNAL ELECTROMAGNETIC FIELD IN CURVED SPACETIME}

\author{YU. I. SHIL'NOV}

\address{CSIC, Institut D'Estudis Espacials De Catalunya,
Ed. Nexus-104,\\ Gran Capita 2-4, 08034, Barcelona, Espa\~na\\
E-mail: visit2@ieec.fcr.es\\
and\\
 Department of Theoretical Physics, Faculty of Physics,\\
Kharkov State University, Svobody Sq. 4,\\
 310077, Kharkov, Ukraine} 

\maketitle\abstracts{ An investigation  of the Nambu-Jona-Lasino model
with external constant electromagnetic
and weak gravitational fields  is carried out in three- and
four-dimensional
spacetimes. The effective potential of the
composite bifermionic fields is calculated keeping terms linear
in the curvature,
 while the  electromagnetic field effect is treated exactly
by means of the proper-time formalism.
Numerical simulations of the  dynamical
symmetry breaking phenomenon accompanied by some phase transitions 
are presented.}

1. Spontaneous symmetry breaking phenomenon is the absolutely necessary
part of modern quantum field theory. It has very strong experimental
basis and is usually  incorporated in any new model.

2. One of the most attractive opportunities to involve the spontaneous
generation of  dimensional values   is the dynamical symmetry
breaking mechanism when  masses, effective coupling
constants and so on are expressed through the vacuum expectation value
of bifermionic  composite field
$\langle {\overline \psi}\psi \rangle$
instead of some hypothecal Higgs scalar field
$\langle\varphi\rangle$.

3. Dynamical symmetry breaking has been applied successfully to describe
the overcritical behaviour of quantum electrodynamics, top quark
condensate
mechanism of mass generation in Weinberg- Salam model,
technicolor models and especially to investigate
nonpertubatively the composite fields generation in four-fermionic models
that of Nambu- Jona- Lasinio one.

4. Recently this model has been studied in external electromagnetical
field.
The essential role of this field in the dynamical symmetry breaking
realisation has been proved.
Furthemore  it has been paid a great attention to the dynamical
symmetry breaking phenomenon in curved spacetime.

5. It seems to be important to build up the realistic scenario of early 
Universe. However it has been shown early Universe should contain
a large primodial magnetic field and have a huge electrical conductivity.

It makes us consider the dynamical symmetry breaking and dynamical fermion
mass generation with the presence of both electromagnetic and
gravitational fields.

We consider  Nambu -- Jona -- Lasinio model in external constant
electromagnetic field treated nonpertubatively in Schwinger propertime
formalism. The linear -- curvature corrections for the effective
potential of composite bifermionic field are  calculated and the phase
structure of the model is investigated.

Both 4D and 3D cases are disussed.
The phase transitions accompanied dynamical symmetry breaking on the    
spacetime curvature and electromagnetic field strength values are
described numerically.
\bigskip
 Let us discuss now the Nambu- Jona- Lasinio model in the
arbitrary dimension curved space- time with the following action:
$$
S=\!\int d^d x \sqrt{-g} \{i \overline{\psi}\gamma^\mu (x)D_\mu \psi +
{\lambda \over 2N} \biggl[ (\overline{\psi}\psi)^2+
(\overline{\psi} i \gamma_5 \psi)^2 \biggr] \},
\eqno(1)
$$
where the covariant derivative $D_{\mu}$ includes the electromagnetic
potential $A_{\mu}$,
local Dirac matrices $\gamma_\mu (x)$ are expressed through the usual
flat one $\gamma_a$ and tetrads $e^a_\mu$,
$\sigma_{ab}={1\over 4 }[\gamma_a,\gamma_b]$,
$\omega^{ab}{}_{\!\mu}$ is a spin- connection and N is the number of
bispinor fields $\psi_a$. Spinor
representation dimension is supposed to be four. Greek and Latin indices
correspond to the curved and flat tangent spacetimes.
  
Introducing the auxiliary fields
$\sigma=-{\lambda \over N }(\overline{\psi} \psi ),
\pi=-{\lambda\over N}\overline{\psi} i \gamma_5 \psi$ ,
we can rewrite the action as:
$$
S=\int d^d x \sqrt{-g} \{ i\overline{\psi}\gamma^\mu D_\mu \psi -
{N \over 2\lambda}(\sigma^2+\pi^2)-
\overline{\psi}(\sigma+i\pi\gamma_5)\psi\}.
\eqno(2)
$$

Then the effective potential is given by:
$$
V_{eff}={\sigma^2 \over 2\lambda }+i \Sp \ln \langle x| [ i\gamma^\mu
(x)D_\mu - \sigma] |x \rangle
\eqno(3)
$$
  
By means of the usual Green function, which obeys the equation
$$
(i \gamma^\mu D_\mu-\sigma)_x G(x,x',\sigma)=\delta(x-x'),
\eqno(4)
$$
we obtain the following formula
$$
V_{eff}'(\sigma)={ \sigma \over \lambda }-i \Sp G(x,x,\sigma)
\eqno(5)
$$

To calculate the linear curvature corrections the local momentum
expansion formalism is the most convinient one. Then in the
special Riemannian normal coordinate framework
$$
g_{\mu\nu}(x)=\eta_{\mu\nu}-{1\over 3 }
R_{\mu\rho\sigma\nu}y^\rho y^\sigma
\eqno(6)
$$
and corresponding formulae for the others values with $y=x-x'$.
Then we suppose that:
$$
G(x, x', \sigma)=
\Phi (x, x')\left[ \tilde {G}_0 (x - x', \sigma)+
\tilde {G}_1 (x - x', \sigma) \dots \right],
\eqno(7)
$$
where $\tilde G_n \sim R^n $,
$\Phi(x,x')=\exp  [ie\int^x_{x'} A^\mu (x'') dx''_{\mu}]$
and therefore our basic expression for the Green function
 $\tilde{G_1}$ in a constant curvature space-time of arbitrary
dimension $d$  is the following:   
$$
\tilde{G_1}(0,\sigma)=-\frac{iR}{12d(d-1)}\int dz G_{00}(-z,
\sigma)
\biggl[ 2\!\not{\!z}z^\mu\partial_\mu \tilde{G_0}(z,\sigma)
$$
$$
-2z^2\gamma^\mu \partial_\mu \tilde{G_0}(z,\sigma)+3(d-1)\!\not{\!z}
\tilde{G_0}(z,\sigma)],
\eqno(8)
$$
where $G_{00}(z, \sigma)$ is the free Green function.
This expression can be substituted into Eq. (5) directly, because
$\Phi(x, x)=1$.

Sustituting an exact flat spacetime Green function $G_{0}(z,\sigma)$
of fermions in external
electromagnetic field  into this formula
after some algebra we have evident expression for the
effective potential  with the linear-~curvature accuracy   
in the constant curvature  spacetime.
  
For 3D spacetime effective potential in the
 constant magnetic field case is given by:
$$
V_{eff}(\sigma)={\sigma^2\over 2\lambda}+{1\over 4\pi^{3/2}}
\int_{1/\Lambda^2}^\infty {ds\over s^{5/2}}\exp[-s\sigma^2]\tau
\coth\tau-
$$
$$
{R\over 144\pi^{3/2}}\int_{1/\Lambda^2}^\infty
\int_{1/\Lambda^2}^\infty
\frac{ds dt}{(t+s)^{5/2}(1+\kappa
\coth\tau)^2}\exp[-(t+s)\sigma^2]\times
$$
$$
\biggl[2\kappa(\kappa+\tau)+
(9\tau+5\kappa)\coth\tau+\kappa(\tau-3\kappa)\coth^2\tau\biggr]
\eqno(9)
$$
and for 4D case:

$$
V_{eff}(\sigma)={\sigma^2\over 2\lambda}+{1\over 8\pi^2}
\int_{1/\Lambda^2}^\infty {ds\over s^3} \exp(-s\sigma^2)\tau
\coth\tau-
$$
$$
{R\over 192\pi^2}\int_{1/\Lambda^2}^\infty \int_{1/\Lambda^2}^\infty
\frac{ds dt}{(t+s)^3(1+\kappa \coth\tau)^2}\exp
[-(t+s)\sigma^2]\times
$$
$$
[\kappa(\kappa+\tau)+
2(\kappa+3\tau)\coth\tau+2\kappa(\tau-\kappa)\coth^2\tau],
\eqno(10)
$$
where $\tau=eBs, \kappa= eBt$.
Typical behaviour of effective potential in this case   
is shown on the Fig.1.
\begin{figure}[tn]
    \begin{center}
    \input{vR3_eepic.tex}
    (a) $V_{eff}$ for $R=3\mu^2$.
    \end{center}
    \begin{center}
    \input{vB5_eepic.tex}
    (b) $V_{eff}$ for $eB =\mu^{2}/2$.
    \end{center}
\vglue 1ex
\caption{The behavior of the effective potential $V_{eff}$
is shown with the varying $B$ or $R$ for fixed $\lambda (=1/2.5)$
and fixed $\Lambda (=10\mu)$ for 4D spacetime.
Second-~order phase transition takes place on both magnetic field
strength and spacetime curvature.}
\end{figure}
    
A proper-time representation of  fermion Green function in
external constant electrical field contains after Wick rotation
$\cot(eEs)$ in the contrast of magnetic field case with nonsingular
$\coth(eBs)$. Therefore imaginary part of effective potential caused by
residue contribution appears. It means that particle creation takes place
and our vacuum is unstable.
    
The solution of this problem lies out of our present investigation but
the simplest possibility seems to consider a comparably small electrical
field strength values which provide an exponentially depressed particle
creation velocity.

3D four- fermions models have been shown to be  renormalizable in the
leading large-N order. Furthemore the external electromagnetic and
gravitational field don't influence with  the renormalization procedure
 It caused by the
fact that the local feature of renormalizability can't be spoiled
by the weak curvature of global spacetime or external electrical field.
Formally it can be proved by the integrand leading terms calculations for
the $s\to 0, t\to 0$ limit which is the only  essential to determine
the  UV-divergences of effective potential.
    
So  after coupling constant renormalization
$$
\frac{1}{\lambda_R}=\frac{1}{\lambda}-\frac{\Lambda}{3\pi^2}
\eqno(11)
$$
we have for renormalized effective potential the following expression:
$$
V_{eff, R}^{(3D)}(\sigma)={\sigma^2\over 2\lambda_R}-
{(2ieE)^{3/2} \over 4\pi}
\biggl[ 2 \zeta (-{1\over2},{\sigma^2 \over 2ieE})-
\biggl({\sigma^2\over 2ieE}\biggr)^{1/2} \biggr]+
$$
$$
{R\sigma \over 24\pi}+{iR(eE)^{1/6} \over 2\pi^2 3^{7/3}}
\exp(-\pi{\sigma^2 \over eE})\Gamma(\frac{2}{3})\sigma^{2/3}.
\eqno(12)
$$
However 4D four- fermionic models are not renormalizable. Thus we have to
make a trick and consider "almost" 4D situation with $d=4-2\epsilon$ where
renormalizability feature exists. Then effective potential is given by:
$$
V^{(4D)}_{eff}(\sigma)={\sigma^2 \over 2 \lambda} -  
\frac{(eE)^2}{8\pi^2}\Gamma(-1 + \epsilon)
\left[4\zeta(-1 + \epsilon, -i{\sigma^2\over 2eE}) +
 i{\sigma^2 \over eE}\right]-
$$
$$
{R \sigma^2 \over 96\pi^2}\Gamma(-1+\epsilon)+
{iR(eE)^{2/3} \over 48\pi^3 3^{1/3}}
\exp(-\pi{\sigma^2 \over eE})\Gamma(\frac{2}{3})\sigma^{2/3}
\eqno(13)
$$

The results of numerical simulation are presented on the Fig.2-~Fig.3.

\begin{figure}[t]
\psfig{figure=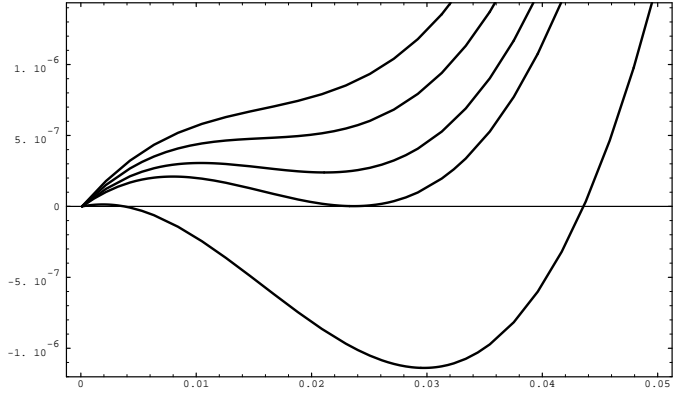,height=4.5in}
\caption{Behaviour in 3D of $\Re V_{eff, R}/\mu^3$
as a function
of $\sigma/\mu$  for
fixed $eE/\mu^2= 0.00005$ and  $ \lambda\mu= -100$.
From above to below,  the curves in the  plot correspond to
the following values of
$R/\mu^2=0.006; 0.005; 0.004; 0.0032; 0$, respectively.
The critical values, defined as usual, are given by:
$R_{c1}/\mu^2=0.005$; $R_{c}/\mu^2=0.0032$;
 $R_{c2}/\mu^2=0$. First-~order phase transition takes place.}
\end{figure}
\begin{figure}[t]
\psfig{figure=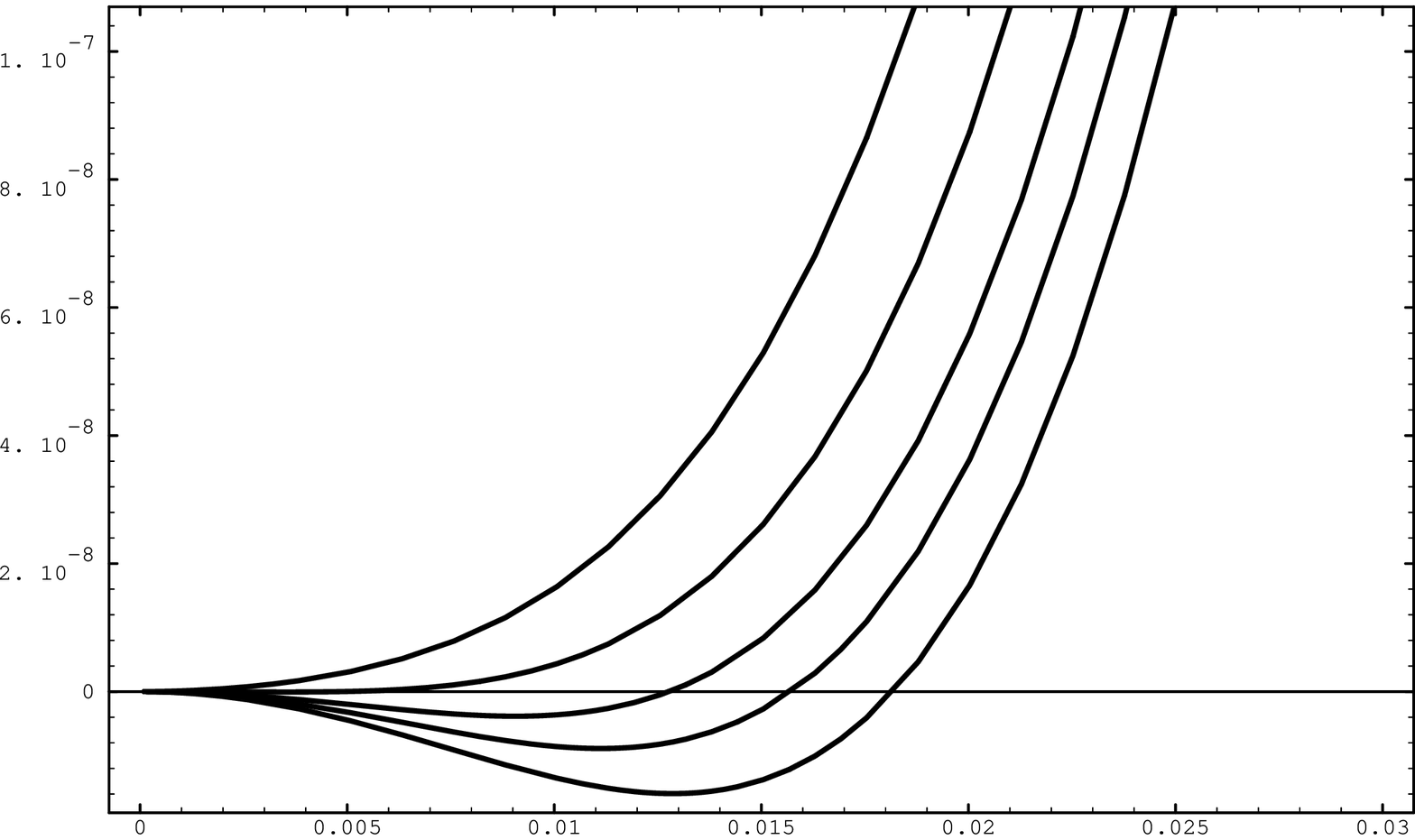,height=4.5in}
\caption{Behaviour of
$\Re V_{eff}/\mu^{(4-2\epsilon)}$
 in a $(4-2\epsilon)$-dimensional spacetime,
as a function
of $\sigma/mu$, depicted for
$R/\mu^2= 0.002$, $\epsilon = 0.005$
 and fixed $\lambda\mu= -1000$.
From above, the curves in the plot correspond to
the following values of the electric field strength:
$eE/\mu^2 = 0.004; 0.0028; 0.02; 0.015; 0.001$, respectively.  
The critical value is reached at
$eE_{c}/\mu^2=0.0028$. Second-~order phase transition occurs.}
\end{figure}

1. We have studied the phase structure of Nambu-Jona-Lasinio model
 in curved spacetime with external constant  electrical and magnetic
fields.

2. In three dimensions
first-order phase transition takes place both on electrical field
strength and on spacetime curvature. Magnetic field induced second-order
phase transition.

3. In four- dimensional case
the typical second-order one  occurs on both these external paremeters.

4. Our approximation seems to be correct because all of the critical
values are very small.

5. Positive spacetime curvature tries to restore chiral symmetry
as well as external electrical field meanwhile  magnetic
field breaks symmetry for any  finite strength value.

\bigskip

The results presented in this talk have been obtained in collaboration
with V. V. Chitov, E. Elizalde, T. Inagaki and S. D. Odintsov.
The work   was supported in part by Ministerio de
Educaci\'on y Cultura (Spain), grant SB96-A04620572. I'd like
also to express
my  deep gratitude to A. Letwin and R. Patov for their kind support
and finally to the organizers of this really charming Workshop.
\end{document}

%% file: vR3_eepic.tex
\setlength{\unitlength}{0.240900pt}
\begin{picture}(900,809)(0,0)
\thicklines \path(220,174)(240,174)
\thicklines \path(836,174)(816,174)
\put(198,174){\makebox(0,0)[r]{-0.008}}
\thicklines \path(220,297)(240,297)
\thicklines \path(836,297)(816,297)
\put(198,297){\makebox(0,0)[r]{-0.006}}
\thicklines \path(220,419)(240,419)
\thicklines \path(836,419)(816,419)
\put(198,419){\makebox(0,0)[r]{-0.004}}
\thicklines \path(220,541)(240,541)
\thicklines \path(836,541)(816,541)
\put(198,541){\makebox(0,0)[r]{-0.002}}
\thicklines \path(220,664)(240,664)
\thicklines \path(836,664)(816,664)
\put(198,664){\makebox(0,0)[r]{0}}
\thicklines \path(220,786)(240,786)
\thicklines \path(836,786)(816,786)
\put(198,786){\makebox(0,0)[r]{0.002}}
\thicklines \path(220,113)(220,133)
\thicklines \path(220,786)(220,766)
\put(220,68){\makebox(0,0){0}}
\thicklines \path(357,113)(357,133)
\thicklines \path(357,786)(357,766)
\put(357,68){\makebox(0,0){0.2}}
\thicklines \path(494,113)(494,133)
\thicklines \path(494,786)(494,766)
\put(494,68){\makebox(0,0){0.4}}
\thicklines \path(631,113)(631,133)
\thicklines \path(631,786)(631,766)
\put(631,68){\makebox(0,0){0.6}}
\thicklines \path(768,113)(768,133)
\thicklines \path(768,786)(768,766)
\put(768,68){\makebox(0,0){0.8}}
\thicklines \path(220,113)(836,113)(836,786)(220,786)(220,113)
\put(45,899){\makebox(0,0)[l]{\shortstack{$V_{eff}(\sigma)/\mu^4$}}}
\put(528,23){\makebox(0,0){$\sigma/\mu$}}
\put(268,725){\makebox(0,0)[l]{{\small $B=0$}}}
\put(384,664){\makebox(0,0)[l]{{\small $eB=\mu^2/2$}}}
\put(453,572){\makebox(0,0)[l]{{\small $eB=\mu^2$}}}
\put(521,388){\makebox(0,0)[l]{{\small $eB=3\mu^2/2$}}}
\put(350,174){\makebox(0,0)[l]{{\small $eB=2\mu^2$}}}
\thinlines \path(220,664)(220,664)(221,664)(242,665)(263,666)(284,669)(305,671)(326,674)(348,678)(369,682)(390,687)(411,693)(432,701)(453,710)(475,722)(496,737)(517,755)(538,777)(545,786)
\thinlines \path(220,664)(220,664)(221,664)(245,669)(270,674)(294,678)(318,682)(343,685)(367,688)(392,691)(416,695)(440,701)(465,709)(489,720)(514,736)(538,756)(562,782)(565,786)
\thinlines \path(220,664)(220,664)(221,664)(235,666)(243,666)(246,667)(248,667)(249,667)(250,667)(251,667)(252,667)(254,667)(257,667)(259,667)(260,667)(261,667)(262,667)(263,667)(265,667)(268,667)(272,667)(279,666)(309,662)(338,654)(367,645)(396,634)(426,625)(455,618)(470,616)(477,615)(481,615)(484,615)(486,615)(487,615)(488,615)(489,615)(490,615)(492,615)(494,615)(495,615)(499,615)(514,617)(543,626)(572,643)(602,671)(631,712)(660,766)(668,786)
\thinlines \path(220,664)(220,664)(221,664)(228,664)(230,664)(232,665)(233,665)(234,665)(234,665)(235,665)(236,665)(237,665)(239,664)(243,664)(250,662)(279,653)(309,635)(338,613)(367,585)(396,556)(426,526)(455,498)(484,472)(514,451)(543,437)(558,433)(565,432)(569,432)(570,432)(572,432)(573,432)(574,432)(575,432)(576,432)(578,432)(580,432)(587,433)(602,436)(631,453)(660,484)(689,532)(719,597)(748,683)(776,786)
\thinlines \path(220,664)(220,664)(221,664)(224,664)(225,664)(226,664)(227,664)(228,664)(229,664)(230,664)(232,664)(235,663)(239,663)(243,661)(250,658)(279,637)(309,605)(338,565)(367,517)(396,465)(426,411)(455,357)(484,305)(514,257)(543,215)(572,181)(602,156)(616,149)(624,146)(631,144)(635,144)(636,144)(638,144)(639,144)(640,143)(641,143)(642,143)(644,144)(646,144)(653,145)(660,147)(689,165)(719,202)(748,259)(777,339)(807,444)(836,577)
\end{picture}

%% file: vB5_eepic.tex
\setlength{\unitlength}{0.240900pt}
\begin{picture}(900,809)(0,0)
\thicklines \path(220,174)(240,174)
\thicklines \path(836,174)(816,174)
\put(198,174){\makebox(0,0)[r]{-0.008}}
\thicklines \path(220,297)(240,297)
\thicklines \path(836,297)(816,297)
\put(198,297){\makebox(0,0)[r]{-0.006}}
\thicklines \path(220,419)(240,419)
\thicklines \path(836,419)(816,419)
\put(198,419){\makebox(0,0)[r]{-0.004}}
\thicklines \path(220,541)(240,541)
\thicklines \path(836,541)(816,541)
\put(198,541){\makebox(0,0)[r]{-0.002}}
\thicklines \path(220,664)(240,664)
\thicklines \path(836,664)(816,664)
\put(198,664){\makebox(0,0)[r]{0}}
\thicklines \path(220,786)(240,786)
\thicklines \path(836,786)(816,786)
\put(198,786){\makebox(0,0)[r]{0.002}}
\thicklines \path(220,113)(220,133)
\thicklines \path(220,786)(220,766)
\put(220,68){\makebox(0,0){0}}
\thicklines \path(357,113)(357,133)
\thicklines \path(357,786)(357,766)
\put(357,68){\makebox(0,0){0.2}}
\thicklines \path(494,113)(494,133)
\thicklines \path(494,786)(494,766)
\put(494,68){\makebox(0,0){0.4}}
\thicklines \path(631,113)(631,133)
\thicklines \path(631,786)(631,766)
\put(631,68){\makebox(0,0){0.6}}
\thicklines \path(768,113)(768,133)
\thicklines \path(768,786)(768,766)
\put(768,68){\makebox(0,0){0.8}}
\thicklines \path(220,113)(836,113)(836,786)(220,786)(220,113)
\put(45,899){\makebox(0,0)[l]{\shortstack{$V_{eff}(\sigma)/\mu^4$}}}
\put(528,23){\makebox(0,0){$\sigma/\mu$}}
\put(316,737){\makebox(0,0)[l]{{\small $R=3\mu^2$}}}
\put(453,609){\makebox(0,0)[l]{{\small $R=2\mu^2$}}}
\put(494,529){\makebox(0,0)[l]{{\small $R=\mu^2$}}}
\put(514,437){\makebox(0,0)[l]{{\small $R=0$}}}
\put(542,327){\makebox(0,0)[l]{{\small $R=-\mu^2$}}}
\put(542,180){\makebox(0,0)[l]{{\small $R=-2\mu^2$}}}
\thinlines \path(220,664)(220,664)(221,664)(245,669)(270,674)(294,678)(318,682)(343,685)(367,688)(392,691)(416,695)(440,701)(465,709)(489,720)(514,736)(538,756)(562,782)(565,786)
\thinlines \path(220,664)(220,664)(221,664)(231,665)(242,665)(247,666)(250,666)(252,666)(255,666)(256,666)(257,666)(258,666)(258,666)(259,666)(260,666)(260,666)(263,666)(274,666)(284,665)(305,663)(326,660)(348,656)(369,652)(390,648)(411,644)(432,641)(443,641)(448,640)(451,640)(453,640)(455,640)(455,640)(456,640)(457,640)(457,640)(459,640)(461,640)(464,640)(475,641)(496,644)(517,650)(538,660)(559,674)(580,694)(601,718)(623,749)(644,786)
\thinlines \path(220,664)(220,664)(221,664)(224,664)(224,664)(225,664)(226,664)(227,664)(228,664)(228,664)(230,664)(233,664)(239,663)(245,663)(270,658)(294,650)(318,640)(343,629)(367,616)(392,604)(416,592)(440,581)(465,572)(489,565)(501,563)(508,563)(511,563)(512,563)(514,563)(514,563)(515,563)(516,563)(517,563)(517,563)(518,563)(520,563)(526,563)(538,565)(562,572)(587,586)(611,607)(636,636)(660,676)(685,726)(708,786)
\thinlines \path(220,664)(220,664)(221,664)(247,659)(273,648)(299,633)(325,614)(351,594)(377,573)(403,551)(429,529)(455,510)(481,492)(507,479)(533,470)(540,468)(546,467)(549,467)(553,467)(554,467)(556,467)(557,467)(558,467)(558,467)(559,467)(561,467)(562,467)(562,467)(566,467)(572,467)(585,470)(611,482)(637,503)(663,536)(689,580)(715,638)(742,710)(764,786)
\thinlines \path(220,664)(220,664)(221,664)(250,654)(279,634)(309,608)(338,578)(367,545)(396,511)(426,477)(455,444)(484,415)(514,389)(543,370)(558,363)(572,358)(580,357)(583,356)(587,356)(589,356)(591,356)(592,355)(592,355)(593,355)(594,355)(595,355)(596,355)(598,356)(602,356)(616,359)(631,364)(660,386)(689,423)(719,476)(748,549)(777,642)(807,760)(812,786)
\thinlines \path(220,664)(220,664)(221,664)(250,650)(279,623)(309,590)(338,551)(367,509)(396,466)(426,422)(455,379)(484,339)(514,303)(543,272)(572,249)(587,241)(602,235)(609,233)(616,232)(620,232)(622,231)(623,231)(624,231)(624,231)(625,231)(627,231)(631,232)(638,233)(646,235)(660,241)(689,265)(719,306)(748,366)(777,447)(807,551)(836,680)
\end{picture}